# GravityCam: Wide-field Imaging Surveys in the Visible from the Ground.

C. Mackay,[1][*] M. Dominik,[2] I.A. Steele,[3] C. Snodgrass,[4] U.G. Jørgensen,[5] J. Skottfelt,[4] K. Stefanov,[4] B. Carry,[6] F. Braga-Ribas,[7] A. Doressoundiram,[8] V.D. Ivanov,[9;10] P. Gandhi,[11] D.F. Evans,[12] M. Hundertmark,[13] S. Serjeant,[4] S. Ortolani[14;15]

[1]*Institute of Astronomy, University of Cambridge, Cambridge CB3 0HA*
[2]*Centre for Exoplanet Science, SUPA School of Physics & Astronomy, University of St Andrews, North Haugh, St Andrews, KY16 9SS*
[3]*Astrophysics Research Institute, Liverpool John Moores University, Liverpool CH41 1LD*
[4]*School of Physical Sciences, The Open University, Milton Keynes, MK7 6AA*
[5]*Niels Bohr Institute & Centre for Star and Planet Formation, University of Copenhagen, Øster Voldgade 5, 1350 Copenhagen, Denmark*
[6]*Universite Cote d'Azur, Observatoire de la Cote d'Azur, CNRS, Laboratoire Lagrange, France*
[7]*Federal University of Technology - Parana (UTFPR / DAFIS), Curitiba, Brazil*
[8]*Observatoire de Paris-LESIA, 5 Place Jules Janssen, Meudon Cedex 92195, France*
[9]*European Southern Observatory, Ave. Alonso de Cordova 3107,Vitacura, Santiago, Chile*
[10]*European Southern Observatory, Karl-Schwarzschild-Str. 2, 85748 Garching bei Munchen, Germany*
[11]*Department of Physics & Astronomy, University of Southampton, Highfield, Southampton, SO17 1BJ*
[12]*Astrophysics Group, Keele University, Staffordshire, ST5 5BG*
[13]*Astronomisches Rechen-Institut, Zentrum fur Astronomie der Universitat Heidelberg (ZAH), 69120 Heidelberg, Germany*
[14]*Dipartimento di Fisica e Astronomia, Universita degli Studi di Padova, Vicolo dell'Osservatorio 3, 35122, Padova, Italy*
[15]*Osservatorio Astronomico di Padova, INAF, Vicolo dell'Osservatorio 5, 35122, Padova, Italy*

*email: cdm <at> ast.cam.ac.uk

**Abstract.** GravityCam is a new concept of ground-based imaging instrument capable of delivering significantly sharper images from the ground than is normally possible without adaptive optics. Advances in optical and near infrared imaging technologies allow images to be acquired at high speed without significant noise penalty. Aligning these images before they are combined can yield a 3-5 fold improvement in image resolution. By using arrays of such detectors, survey fields may be as wide as the telescope optics allows. We describe the instrument and detail its application to accelerate greatly the rate of detection of Earth size planets by gravitational microlensing. GravityCam will improve substantially the quality of weak shear studies of dark matter distribution in distant clusters of galaxies. An extensive microlensing survey will also provide a vast dataset for asteroseismology studies, and GravityCam promises to generate a unique data set on the population of the Kuiper belt and possibly the Oort cloud.

## 1 Introduction

GravityCam originated as a project because of two very different astronomical programs each of which was limited substantially by atmospheric seeing, even on the best sites. The two programs were firstly, the detection of weak gravitational shear in distant clusters of galaxies and secondly the detection of large numbers of Earth-sized exoplanets. Our experience of using Lucky Imaging (Baldwin et al., 2008) for a wide range of targets had shown that even with 100% selection, the use of fast wide-field imaging would allow the tip tilt component of atmospheric turbulence to be removed and allow the resolution of images on ground-based telescope to be improved by a factor of between 2.5 and 3.

A more detailed account of the scientific case for GravityCam may be found on the ArXiv archive website (Mackay et al., 2017). In summary, however, the most promising way of detecting the distribution of dark matter at high redshifts is by looking at the way in which the images of distant galaxies are slightly elongated as the light from those galaxies is lensed gravitationally by intervening clusters of galaxies. These intervening clusters themselves contain substantial amounts of dark matter. The difficulty is that the elongations are typically only in the region of 0.1 arcseconds. Observations made from the ground with a resolution of one arc second on a good site only allow these elongations to be measured relatively inaccurately. That in turn forces the use of much nearer galaxies which are much less affected because the light from them has not travelled so far. Improving the angular resolution on ground-based telescope to 0.25-0.3 arcseconds will undoubtedly make the detection of 0.1 arc second elongation much easier.

The other program which will principally concern us in this paper is the business of trying to detect much larger numbers of Earth-sized exoplanets. At present we know of a large number of exoplanets but these are principally massive gas giant planets orbiting much closer to their central star than the Earth is from our own sun. Those giant planets are much easier to detect than planets as small as the earth. Even so they turn out to be very hard to detect.

The consequence of this is that we actually have very poor knowledge of the mass function of the exoplanets in our own Galaxy, particularly at the lower mass end from about five Earth-masses and below. There are indications that there may be a small number of Earth-sized planets around the great majority of stars in our Galaxy. If this is true we should be able to quantify that as it is from the detection of such planets that we truly have the best chance of finding evidence that there might be an environment in which life as we know it might survive.

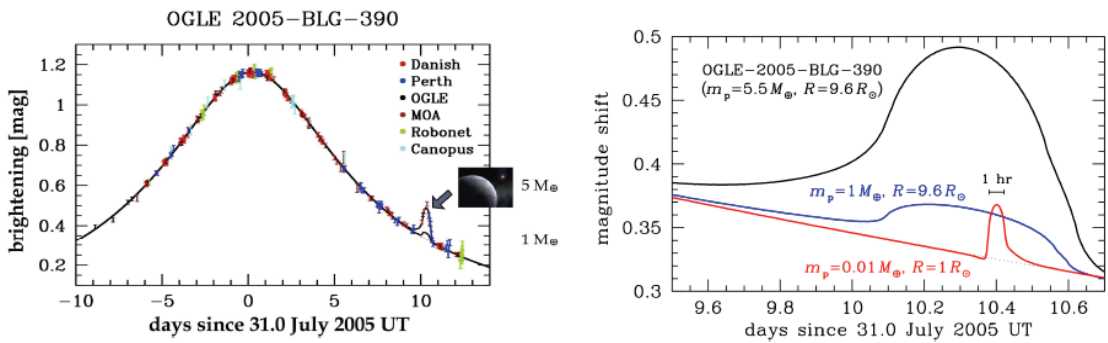

**Figure 1:** These images show on the left the light curve of the star OGLE-2005-BLG-390. This is a star about 9.6 times the diameter of the sun amplifying the brightness of the more distant star by a total of about 1.2 magnitudes (a factor of about 3). The microlensing event can extend over one month or more in total. About 10 days after the maximum of this event a small additional glitch was detected which has been attributed to a planet about five times the mass of the Earth. On the right, we see in black the profile that was actually detected for this exoplanet microlensing event. The width of this occurrence reflects the significantly large size of the lensing star. Had the planet had only one Earth mass but with the same size of central star the profile would have been the middle blue line. Further, had the exoplanet had a mass of only 1% that of the earth, and the star around which planet was orbiting had the same size as our sun the lower profile is what we would predict. This event would only have lasted about one hour in total but the mass of this planet at 1% that of the earth is about the same as that of the moon. No one has ever detected an extrasolar object of that mass before and it demonstrates the power of this method.

The only method that appears to be able to detect earth size planets is by recording the light curve produced by gravitational microlensing of one star by another. When one star moves directly between us and a more distant star, the light from the more distant star is deflected by the gravitational field of the nearer star. This causes the apparent brightness of the more distant star to increase by a significant factor which in some cases is quite small but can also be several thousand times with perfectly aligned lens and source.

In a small number of microlensing events we see deviations from the light curve that we expect. These deviations are due to planets orbiting the lensing star. The existence of planets around the star distorts the gravitational field of the star and has a surprisingly large effect on the light curve that we record on earth. An example of this is shown in figure 1.

Gravitational microlensing of one star by another is a very unusual occurrence. Further, the detection of a planet around a microlensing star is even less likely. If we are to have any chance of detecting these events in reasonable numbers we have to observe very large numbers of stars. The part of our own Galaxy where there are more stars than anywhere else is in the bulge. Indeed the surface density of stars in the bulge of our Galaxy is so great that even at relatively bright levels, typically I~20, the images of stars overlap one another and our capacity to detect these events is badly affected by confusion. In order to overcome this we need to work with a technique to improve the angular resolution on ground-based telescope on a good astronomical site.

Lucky Imaging provides a method for delivering higher resolution on any site. It works simply by taking images of the field of interest at high speed (10-30 Hz frame rate). These are very short exposures essentially to freeze the motion of the star images over the field caused by atmospheric turbulence. Most of the power in atmospheric turbulence is on the largest scales. The largest scale is dominated by tip-tilt distortion. If we simply record a series of images taken at high speed, and then register them so that the reference star used on each frame is precisely aligned then everything else in that frame is also properly aligned. The frame size used must be limited because of tip-tilt anisoplanatism to typically one arc minute.

Not only are the images aligned but we find there is a significant range of intrinsic resolution in each of the images. In some cases the images are extremely sharp indeed, and under the right conditions can be close to being diffraction limited. The smaller the fraction of images that we select by angular resolution the better the image quality obtained can be. If we use 100% of the images we can improve the angular resolution by a factor of 2.5-3. If we then select a smaller subset then the image quality can approach the diffraction limit of the telescope provided the telescope itself is not too large.

Most of the work on Lucky Imaging has been done on moderate size telescopes, up to about 2.5 m diameter. Images obtained by Lucky Imaging with modest selection (10-30%) can have a resolution equivalent to that of the 2.5 m Hubble Space Telescope. A detailed summary of the performance obtained with Lucky Imaging on the NOT 2.5 m telescope on La Palma has been given by Baldwin et al. (2008).

Unfortunately if we try to use this technique on bigger telescopes the chance of getting a sharp images rapidly becomes very low and as the selection percentage needed to deliver the sharpest resolution becomes small. That in turn causes the overall sensitivity of the method to be diminished. However under all circumstances, almost no matter how bad the atmospheric seeing is, using Lucky Imaging on larger telescopes will improve the resolution by this base factor of 2.5-3.

## 2     Detecting Earth-Sized Exoplanets.

The main problem in using gravitational microlensing to detect earth -sized planets is that gravitational microlensing events are quite rare, and microlensing events which involve a planet around the star doing the microlensing are even rarer. To have a successful campaign we need to be able to observe very large numbers of stars to have any hope of accumulating enough significant detections.

The part of our Galaxy which has the highest density of stars is its bulge. The bulge is that part of the galaxy that provides an extended region relatively close to the galactic centre. Most parts of the sky towards the centre of our Galaxy are quite heavily obscured. There are, however, half a dozen windows where the dust of obscuration is relatively small. In those regions the star density is very high.

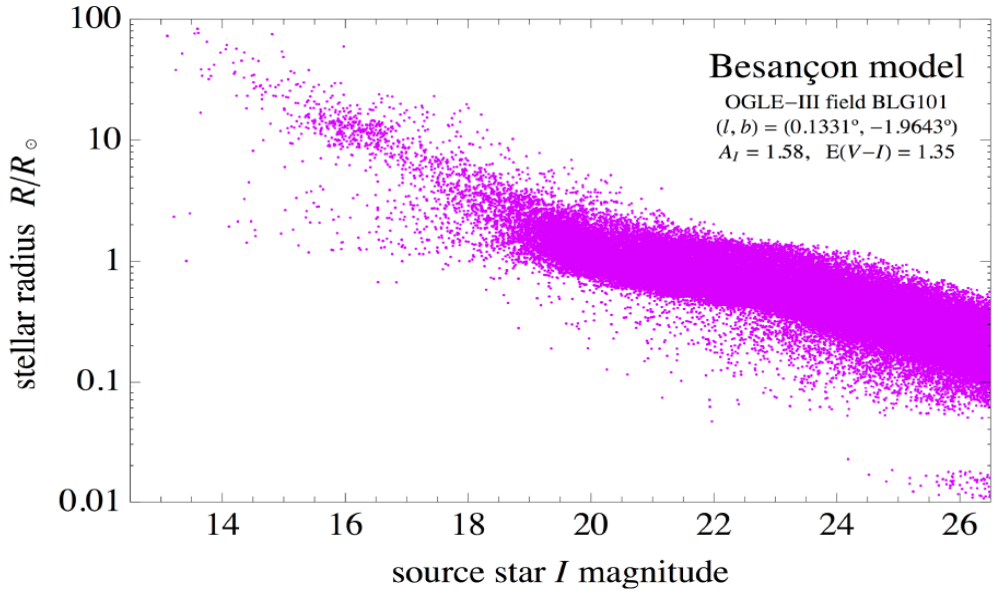

**Figure 1:** the distribution of stars in one field of the Galactic bulge showing the size of the star in units of the size of our Sun against I magnitude. Stars brighter than I ~ 18 are markedly bigger than our and that gives a gravitational microlensing event which is significantly smaller than we would with a main sequence star of the same brightness. The data shown here have been created using the Besancon model of galactic populations (Robin et al, 2003)

If we look at the distribution of stellar radius in terms of the radius of our Sun against the I magnitude (figure 2) we see that the stars brighter than I ~ 18 are much larger. That will generally make it harder to detect gravitational microlensing because the size of the star means the effect of the magnification is diminished. Ideally we want to look at main sequence stars and so we see that from figure 1 we will always want to design to detect stars fainter than I ~ 18.

The star densities in the bulge of the galaxy are very high indeed. In Figure 2, we reproduce an image from the Hubble Space Telescope of part of Baade's window, taken from Holtzmann et al, 1998. The area shown in this picture is about 0.6% of the area of the field.

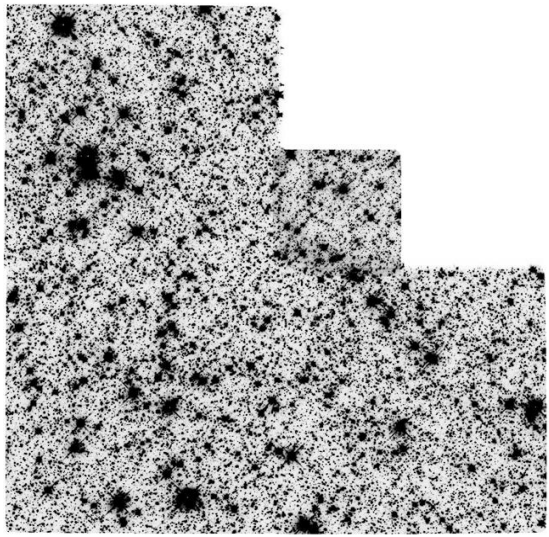

**Figure 2:** The image reproduced here is from the Hubble Space Telescope of a small part (0.6%) of Baade's window. The star densities are very high indeed. ( Holtzman et al., 1998).

Stars with I ~ 18 correspond to an absolute I magnitude of $M_I$ ~ 3. The star densities in the bulge of the galaxy have been examined by Holtzman et al. (1998) and they find that in the range of I = 18-22 the star surface densities are ~ 1/sq arcsec (see Figure 3). If we go another factor of 10 fainter and include the range from I = 18-24.5 then the number of stars is a factor of 10 greater so ~ 10/sq arcsec. This is a very high surface density but unfortunately with conventional ground-based seeing there is little prospect of being able to resolve such high density. Observation of any individual star will be greatly affected by confusion leading to considerable doubt as to the quality of the photometry that might be possible.

Over a 30 x 30 arc minute field of view we would expect to find in the region of 25 million stars. With six fields available in that region it means that an imaging system that can cover much of the field of view would have access to over 150 million stars every night.

The desire to observe this number of objects means that we need to rethink the approach to imaging such fields in order to achieve much higher angular resolution.

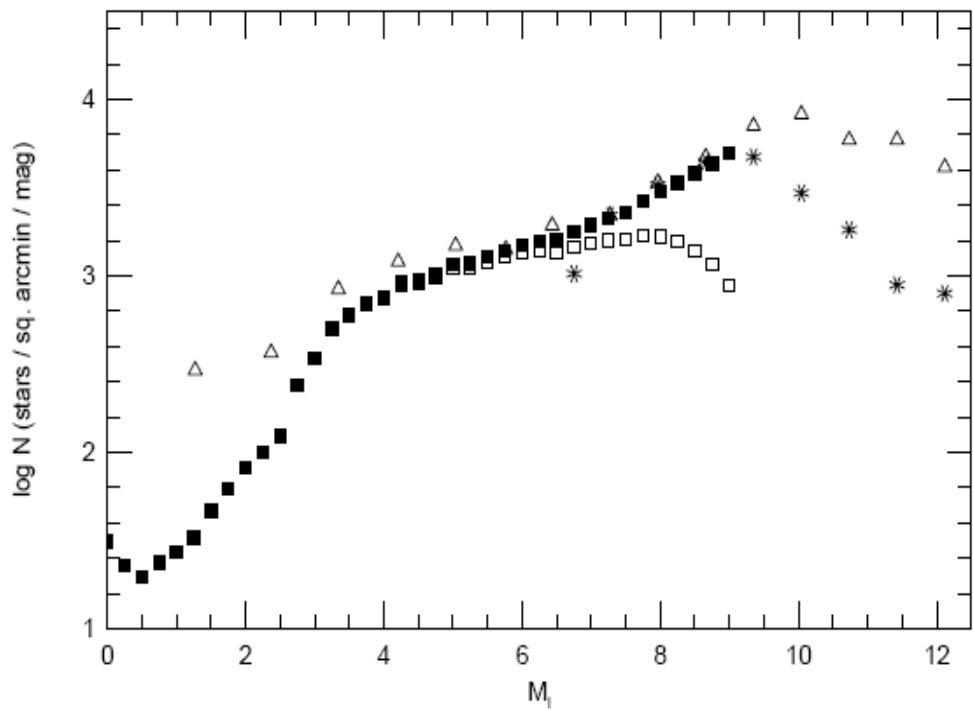

**Figure 3:** The surface densities of stars per square arc minute per magnitude in the bulge of the galaxy. The numbers of stars are very high indeed and we need to work with the integrated number of stars to understand the total star density from the point of view of observability. (Holtzman et al., 1998).

## 3    Lucky Imaging

One of the most straightforward ways to improve the angular resolution of images on a ground-based telescope is to take images rapidly (in the 10-30 Hz range). Using a reference star or group of stars in the field each frame may be registered precisely and summed. This technique is called Lucky Imaging. It was given this name by Fried (1978). When it was originally suggested it was virtually impossible to use the method because of the lack of fast sensitive imaging detectors. Only with the invention of charge coupled devices and particularly those which have internal gain (electron multiplying CCDs, EMCCDs) has this technique come into its own. The technique has been developed by a group in Cambridge for a number of years and the results are best summarised by Baldwin et al. (2008).

The statistics of the atmospheric turbulence guarantee a wide dispersion of image resolution in each frame recorded. Rather than simply some all of them after registration we select the sharpest ones to give a much sharper image. The more restrictive we are the sharper the images but of course the more strict we are in the selection the fewer frames will be used to contribute towards the sum. Fortunately this is something that can be determined after the observations are complete and, for different observing programs, the requirements on resolution may well be different.

The probability of recording a sharp image depends on the telescope size (D) and the characteristic scale size of the turbulent fluctuations in the atmosphere ($r_0$). That cell size, $r_0$, scales directly with the "seeing". On the best optical sites median seeing in the region of 0.6 "-0.8" is achievable. We know from many observing runs summarised by Baldwin et al. (2008) that with a 2.5 m (Hubble -sized) telescope it is possible to take images close to the diffraction limit of the telescope in I-band selecting typically 10-30% of the images to combine into the final image. An example of this is shown in figure 4.

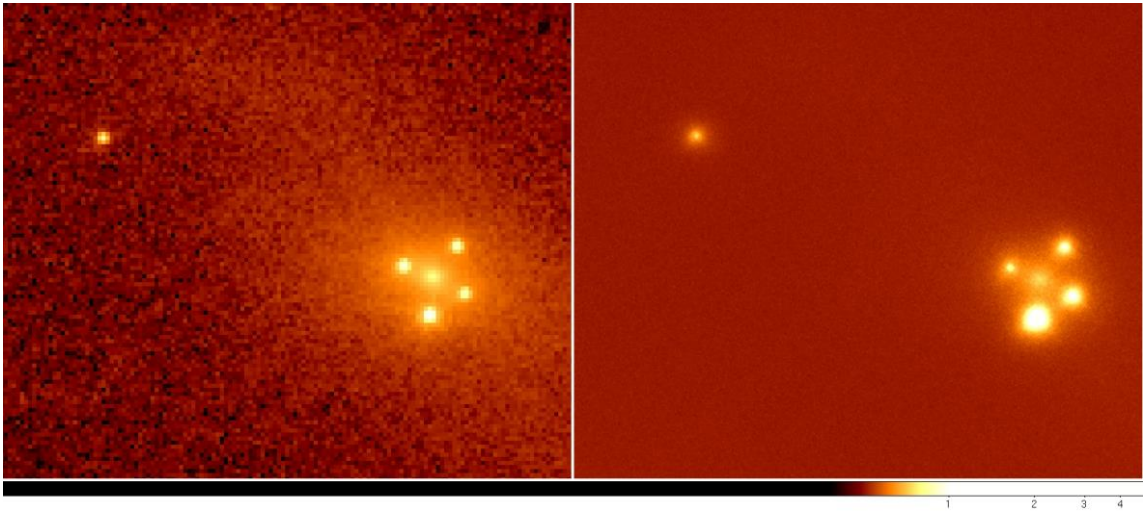

**Figure 4:** The images shown here are of the Einstein cross, where a single distant quasar with the redshift of ~1.7 is gravitationally lensed by a much more nearby Zwicky Galaxy (ZW 2237+030) with the redshift of 0.04. Light from the quasar is bent by the gravitational field of the matter associated with the Galaxy and it produces 4 separate images of the quasar. The quasar itself varies in intensity and the path length from the quasar to the observer depends on which route past the lensing Galaxy it takes. As a consequence, the relative brightness in images taken at a different time can be quite marked. The images here on the left are from the Hubble Space Telescope Advanced Camera for Surveys (ACS) while the image on the right is the Lucky Image taken on the NOT telescope on La Palma in 2009, but also taken through significant amounts of dust. Both telescopes are 2.5 m diameter and the image quality is very similar indeed. The Hubble picture is undersampled which is why the blocky appearance of the image is visible. The lucky camera uses significantly smaller pixels.

The effective seeing on the appearance of images taken in the bulge of the Galaxy is quite dramatic. Figure 5 (from Bennett, 2004) shows images of a microlensing event (MACHO-96-BLG-5)observed with a 0.9 m telescope in one arcsecond seeing (top images) and images from the Hubble Space Telescope where the target star can be seen very easily and clearly separated from its neighbours.

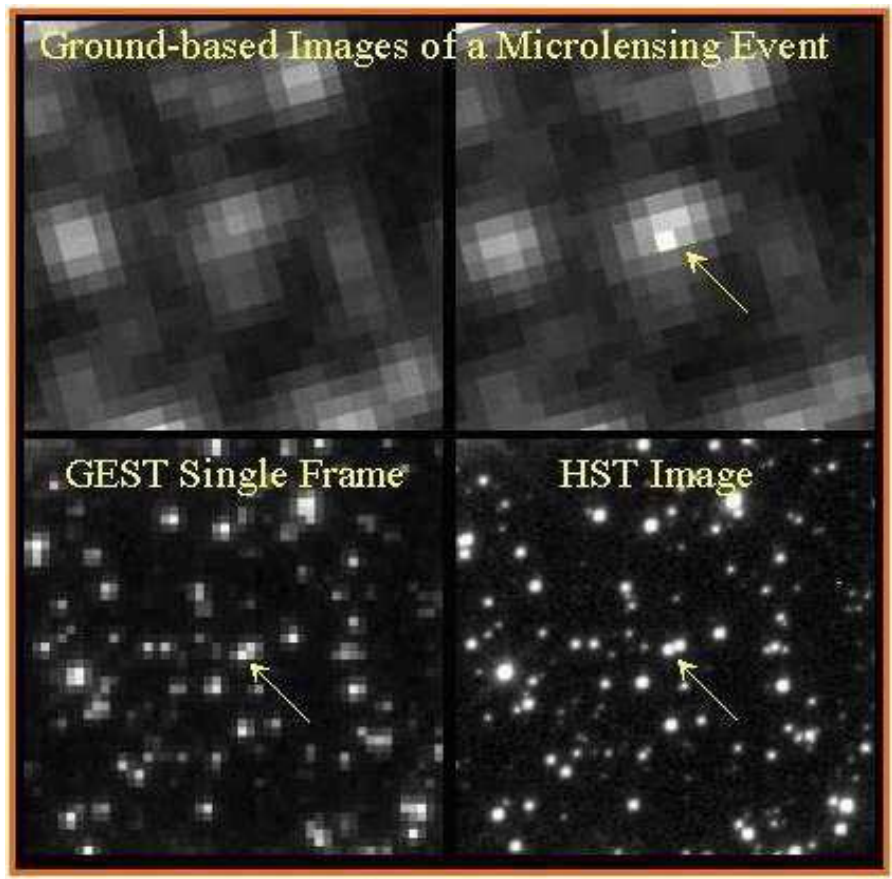

**Figure 5:** There is considerable difference between the quality of the images obtained from the ground under even quite good seeing conditions (the upper images were taken with 0.9 m telescope on Cerro Tololo in Chile with ~ 1 arcsec seeing), while the lower images were taken with the Hubble Space Telescope, both a single frame and a multi frame drizzle image. The lensing star can be seen very clearly and is well separated from its immediate neighbours making it much easier to carry out precision photometry (from Bennett, 2004).

Our experience with Lucky Imaging has shown that under a wide range of atmospheric conditions simply taking images rapidly and registering them before adding gives a resolution a factor of 2.5-3 better than the seeing would normally be expected to deliver. If we select the best 50% of the images then the resolution becomes 3-4 times better.

The choice of large diameter telescopes on which a suitable wide-field camera might be mounted is very limited. Large telescopes generally have very small fields of view while telescopes with big fields of view are generally quite small diameter. If we are indeed to be able to work with rather faint targets stars down to I ~ 24.5 or even fainter we need to be able to use a good sized telescope.

One that would suit our purposes very well in principle is the NTT 3.6 m of the European Southern Observatory site at La Silla in Chile. Chilean sites are particularly attractive because the centre of the Galaxy goes through the zenith at the right time of year. That makes observations much easier to manage and minimises the attenuation that comes from working at larger zenith distances. The NTT has excellent optical performance and was significantly upgraded as part of the route map towards the completion of the VLT at the more Northern site of the ESO at Paranal.

## 4     The GravityCam Instrument.

We have already used a simplified version of GravityCam at the Naysmith focus of the NTT. A picture of our camera mounted on the telescope is shown in figure 6. The Naysmith focus of the NTT has a de-rotator that ensures that North is always up. That is important because one of the key aspects of the instrument is that we will rely on the field centre and orientation to be reliably consistent from night to night and indeed over longer periods.

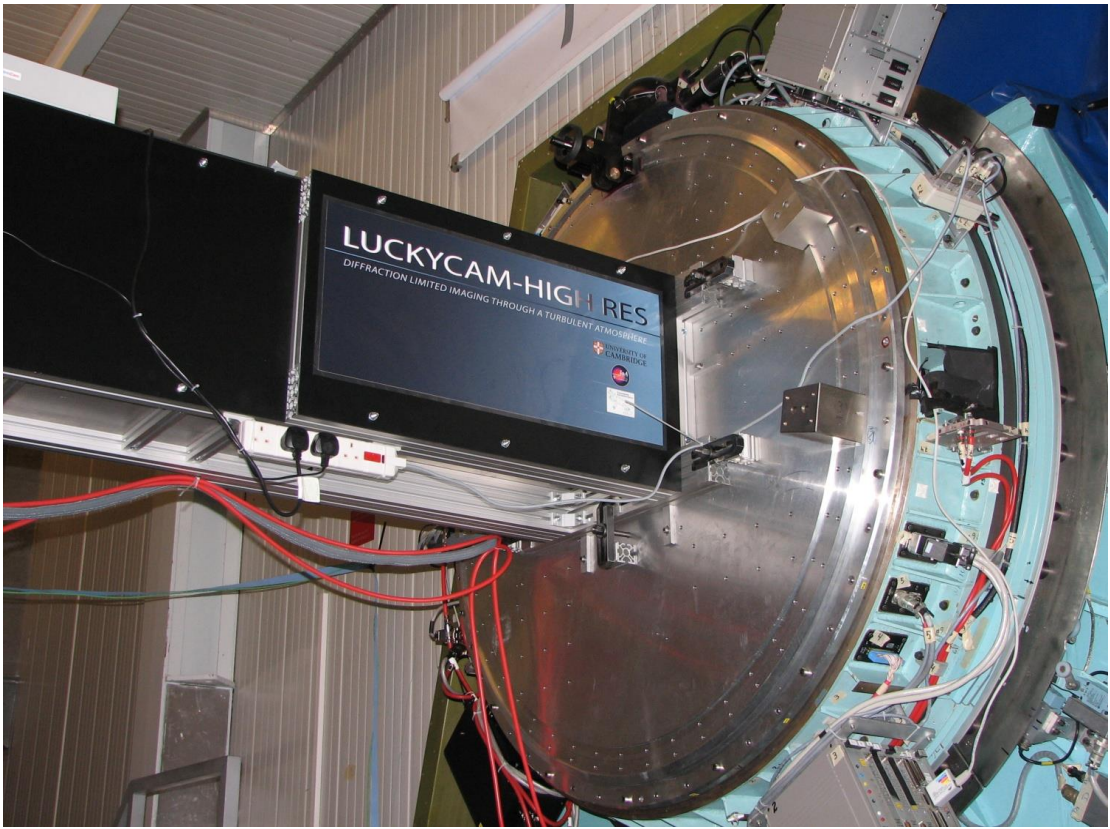

**Figure 6**: Prototype of GravityCam detector mounted on one of the Naysmith platforms at the NTT 3.6 m telescope of the European Southern Observatory in La Silla, Chile. This is one example of the instruments used on a number of telescopes to establish the credentials of the technique on good observing sites such as La Palma in the Canary Islands and La Silla in Chile. The system shown here consisted of a single EMCCD behind a simple crossed prism atmospheric dispersion corrector (ADC) being run in the standard lucky imaging mode. There is considerable space to mount an instrument on the telescope which has extremely high optical quality and is located on a top astronomical site.

La Silla where the NTT is located is an excellent site. Median seeing on the NTT is about 0.75 arcseconds indicating that the median turbulent scale size is $r_0$ ~ 0.24m. We have run extensive simulations based on that seeing and assuming a typical wind speed of ~ 8 m/s. The simulations allow us to generate a phase screen representative of the conditions and then to generate large numbers of frames and from those select different fractions in terms of quality which we define as being the image sharpness or Strehl ratio per frame. Each image is shifted and added in exact coalignment with the other images in that group. We select images that are in the top 1, 5, 10, 20, 30, 50, 100% bins. A typical output from the simulation is shown in figure 7 together with the image for the raw data (lower right) which would be obtained if there was no attempt at shifting and adding the frames.

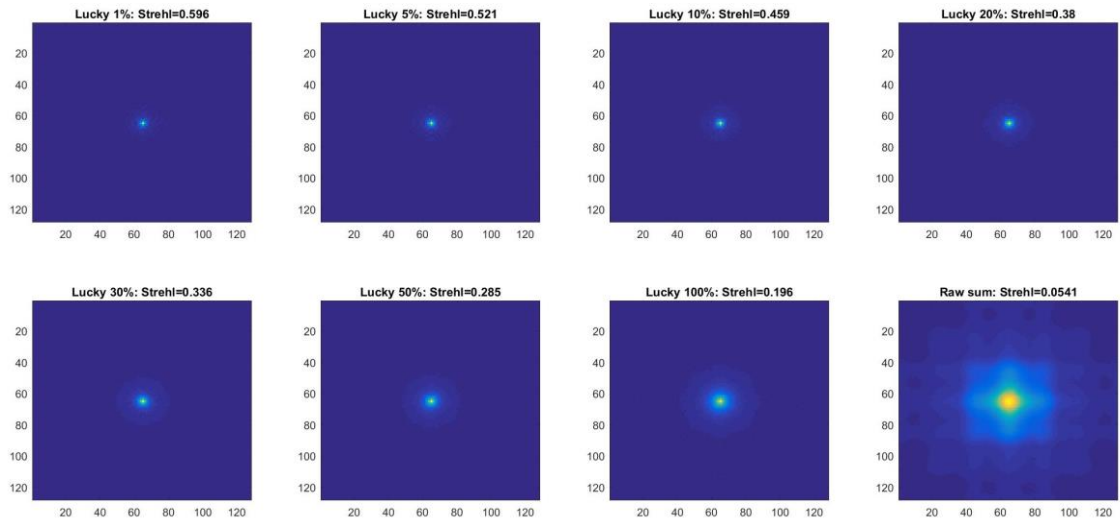

**Figure 7:** The output from a long run of simulated images for GravityCam on the NTT. The seeing was 0.75 arcseconds and the wind velocity 8 m/s. This is a simulation of 1000 images. The image on the bottom right shows what would be obtained with conventional imaging, therefore without any attempt to co-align different images. The other seven frames show the different percentages in each bin and the Strehl ratio which is a measurement of the sharpness of the image produced.

The simulation shown in figure 7 give results that are poorer than with a 2.5 m telescope which can allow near diffraction limited imaging with a higher percentage of frames. However the NTT is on a good site and under the right conditions we should be able to get very good image resolution indeed. Our design for GravityCam does not try to achieve diffraction limited image quality, and sacrifices pixel size for field of view. It is these simulated outputs that have allowed us to select 16 µm pixels which give a scale of 85.76 milliarcseconds per pixel.

That scale would be achieved without any additional reimaging optics. The design of GravityCam is very simple. In the front of the detector enclosure (a vacuum dewar) is an atmospheric dispersion corrector consisting of two crossed prisms. This is essential if we do not wish to compromise the resolution when we observe at significant zenith distances. It is at that point that facilities for introducing broadband filters are included. However many of the programs that are particularly exciting with GravityCam are related to gravitational lensing. In both cases we must remember that the gravitational lensing process is completely achromatic. When signal-to-noise is important it means we can use a very broadband or indeed a simple long pass filter.

The front element of the detector housing will be a weak lens to compensate for the non-flatness of the focal plane. The focal plane from the NTT has a curvature with radius of 1.9 m. Although that is not great it will be significant over the 33 cm diameter of the field of view and without compensation would cause significant defocus at the edge of the field.

Inside the detector housing we will use a close-packed array of CMOS detectors. The original GravityCam concept proposed using photon counting electron multiplying CCDs. They have the ability to run with extremely low read noise and count individual photons. The disadvantage of CCDs generally is that when running at high-speed there is a significant amount of smear caused by the bright stars in the field. As the charge is transferred across the device prior to being read out the light from a bright star gives a trail which will be familiar to those who use CCDs at other telescopes. When running at high frame rates the charge transfer time is a significant fraction of the total integration time and therefore the smearing becomes more noticeable. It is relatively straightforward to correct for it but it still does compromise image quality particularly when working at the limits of the instrument.

There have been major developments with the design of large area CMOS detectors in recent years. Work at the Open University has produced pixel designs which allow the use of deep depletion silicon that can give up to 90% quantum efficiency at 900 nm. These technologies have been used before in CCDs but only now has a method of using them for CMOS detectors been developed. The read noise of CMOS detectors even at high speed is now very good. CMOS devices have been demonstrated with read noise of significantly less than one electron RMS even when running at very high frame rates (Segovia, J. A. et al, private communication).

Large area three edge buttable devices have also been made. An example is shown in figure 8.

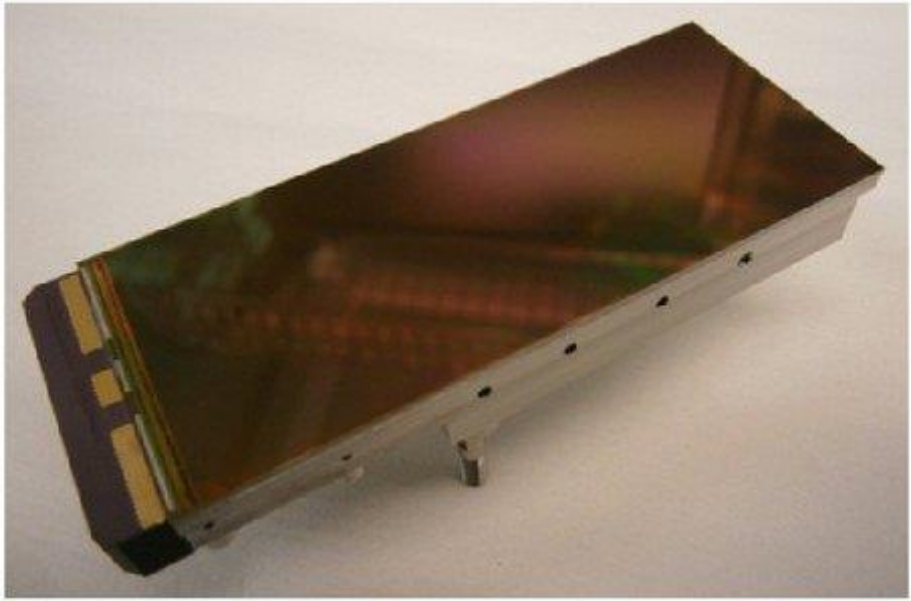

**Figure 8:** Large area CMOS devices are now being manufactured which are three edge buttable. This allows a focal plane to be covered with an array of these detectors with very little loss in sensitive area. The device shown is a CIS113 manufactured by Teledyne-E2V in Chelmsford, UK. This particular device is 1920 x 4608 pixels each of which 16 x 16 µm. The overall detector is approximately 80 mm x 31 mm. This device is back illuminated with analogue output.

It is now possible to integrate much of the readout electronics needed to drive a CMOS detector as well as process the images produced by it. An example of this is in the layout of another device from Teledyne E2V shown in figure 9. This is intended as a very high frame rate device with 75,400 analogue to digital converter channels, with each channel having a programmable gain preamplifier. The digitised data are multiplexed and serialised and taken out in parallel LVDS format.

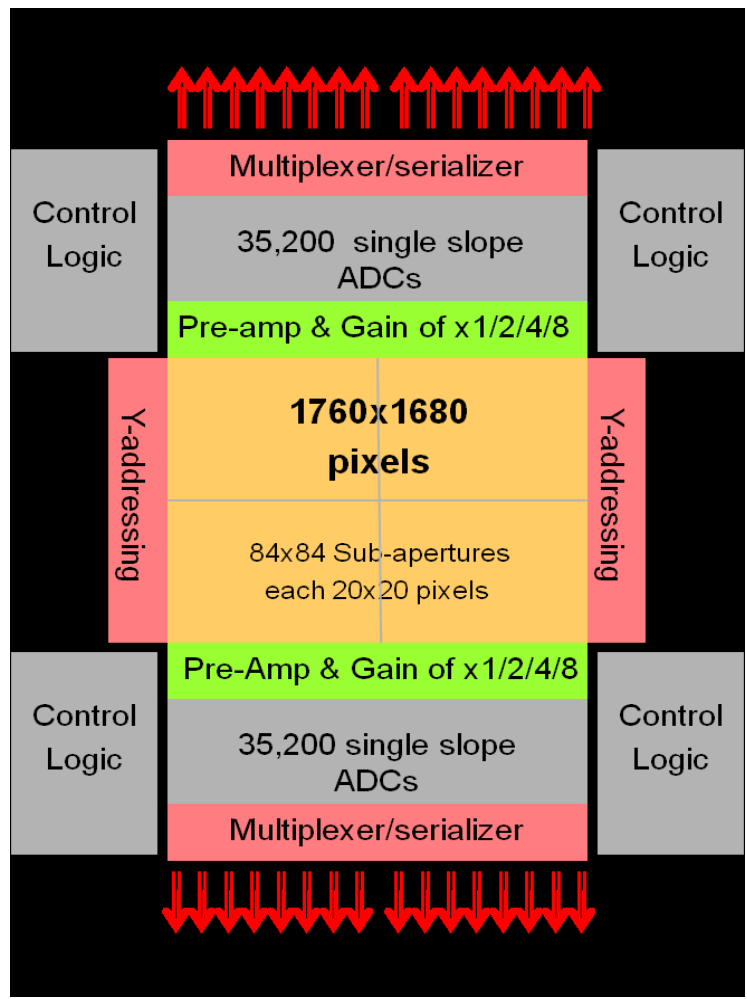

**Figure 9:** An example of the level of integration of control and readout electronics onto the latest generations of CMOS detectors. It is this integration that has a very considerable benefit when building a large array of detectors because the necessity otherwise of providing readout electronics for each detector away from the detector enclosure leads to considerable complexity in the design of the instrument.

The requirements for a CMOS detector for GravityCam can be summarised as follows:

- Must be a three edge buttable device preferably large format. This maximises field coverage.
- Pixels in the size range of 16-20 µm to match the scale of the focal plane of the NTT.
- Frame rate 25 Hz.
- Readout noise less than one electron RMS.
- Full device control electronics, signal processing and digitising on chip.
- Deep depletion silicon in order to give high sensitivity in the 700-900 nm range.

In addition, it is necessary to mount the array of detectors with a very precisely flat focal plane. With 16 µm pixels and a beam at f/8 we need to achieve an overall flatness which is ~15-20 microns.

## 5      Data Management and Processing.

There is no doubt that this will be quite challenging. If we use as a baseline that we would have an array of 36 devices much as the one shown in figure 8 then each of these 9 Mpx devices will each generate about 16 GB per second at 25 Hz frame rate. That gives a total for the entire GravityCam of around 400 TB per night. On no account should any attempt be made to save that quantity of data! It is essential that it is processed substantially in real time. Most likely it will be organised so that each CMOS chip would be attached to its own CPU and at that CPU would contain a graphics processor unit such as one of the Nvidia Volta family and preferably a variant that has direct data access from the camera to the GPU processing board. Exactly how much is done and where it might be done is still uncertain. There are a variety of image processing methods which can be used to improve the image quality and resolution even further. For example the Difference Image Analysis technique describe by Bramich, 2008 has demonstrated quite significant improvements in image resolution. However these techniques do require considerable image processing capabilities but that is something that we anticipate can be managed.

## 6      Predicted Performance for GravityCam.

In order to get some idea of what we might reasonably detect for gravitational microlensing we start by understanding that in fields such as Baade's window, a single pointing of GravityCam will see approximately 25 million stars with I < 24.5. At that limiting magnitude of I = 24.5, stars will produce about 1000 photons in 10 minutes giving a signal-to-noise ratio of ~ 25-30 after sky background correction. At that level it is clear that detector noise is critical.

With an optical depth of about ~$2.4* 10^{-6}$, the estimates from MACHO and OGLE surveys, we predict that we will detect a few new microlensing events every hour. We can survey at least six fields by spending 10 minutes per field in rotation therefore by observing ~150 million stars per night we should detect ~200-500 new microlensing events per night.

The planet detection probabilities are somewhat uncertain. Bennett & Rhie (1996) suggest that we could expect to detect a small number of earth size planets each night. Over a six-month season this becomes a very substantial number of low mass planets detected for the first time.

## 7      Other Applications of GravityCam

There are a number of other applications of GravityCam that are likely to be very rewarding as well. For example the study of dark matter distributions in the universe by looking at the weak gravitational lensing in galaxy clusters. The effect of this weak lensing is to distort the shapes of galaxies slightly, extending them in one direction by about 0.1 arcseconds. With typical ground-based seeing of about one arcseconds it is quite difficult to see such a small extension. With normal ground-based seeing limits such an extension can only be seen with fairly high Galaxy measurement signal-to-noise. That means that most surveys are forced to use relatively nearby bright galaxies where the effect will be less because there is less matter for the light to pass through. With much better seeing the GravityCam can deliver of perhaps 0.3 arcseconds, the detection of a 0.1 arc second extension becomes very much easier. More distant galaxies can be used and therefore the amount of the universe that can be surveyed effectively is greatly increased.

Asteroseismology looks at short-term variations in the light emitted by stars. This allows us to analyse the vibrational spectrum of stars. This gives a considerable amount of information about the internal structure of a star and we have learned a very great deal about the interior of our own sun from the studies. The signal-to-noise on an individual star in the bulge will be much poorer than with solar studies but given that we will have measurements on so many stars night after night for many months it is highly likely that we will get a very good signal-to-noise that lets us establish what the vibration spectrum of different kinds of stars of different spectral type might actually be. That in turn will help us understand the internal structure of these objects.

There is also considerable interest in looking for the occultation of stars by small solar system bodies. Outside the planetary system around the sun there is a region which we believe is made up of material left over from the time that the planets were formed. That material contributes to asteroids and comets that are deflected from the Kuiper belt where they have been orbiting around the sun at great distances so we do have some information about them. However the overall size distribution of objects is completely unknown. The Kuiper belt is thought to extend an angle of 3-5° in the direction of the plane of the ecliptic.

With GravityCam we will be looking at large numbers of stars and in some cases we expect to see some either going out completely for a few milliseconds because an opaque object passes in front of them. The stars are much bigger than probable Kuiper belt objects but the stars are very much further away so a fairly small object can occult to staff at easily. For the first time it should be possible to do a detailed census of the content of the Kuiper belt. Beyond that there is a region known as the Oort cloud. This has been hypothesised as likely to be a cloud of material that almost completely surrounds our Sun. Again we should be able to detect objects should they exist although what actually might be there is very far from known.

We will also be able to gather information about the incidence of planets in orbit round many of the stars. In many cases we will detect at low signal-to-noise dips in the brightness from stars that follow a characteristic pattern that we have seen already in surveys such as that done by the Kepler spacecraft. It is highly unlikely that we will see individual events repeat orbit after orbit unless the occulting planets our large and close to the central star but statistically we should be able to build up a lot of information about the size distribution of exoplanets simply by looking at transit data.

## 8    Conclusions

We have describe GravityCam and shown how Lucky Imaging has the potential to offer a relatively inexpensive way of improving the seeing on ground-based telescopes by a factor of 3-5 (or 9-25 by area) under most conditions on a good site. The design of GravityCam and its capacity to improve the resolution normally experienced on even a good site means that for the first time we will be able to survey large number of stars in high-density fields towards the bulge of the Galaxy.

By observing as many as 15 million stars with I < 24.5 and seeing a significant number of lensing events night after night we should be able to detect several new Earth size planets every night. In addition GravityCam will let us pursue several other important research programs including exoplanet detection, dark matter distribution, Kuiper belt population and others.